# Intrusion Detection with Machine Learning Using Open-Sourced Datasets

**Dr Theodosis Mourouzis[1], Andreas Avgousti[2]**

[1] Electi Consulting LTD; Cyprus International Institute of Management; London Business School; UCL Centre for Blockchain Technologies
[2] Cyprus International Institute of Management, Cyprus

# Introduction

## 1.1 Aim of the Project

No significant research has been conducted so far on Intrusion detection due to data availability since, network traffic within companies is private information and no available logs can be found on the Internet for independent research. This paper aims to answer the question whether open-sourced data, that is usually simulated network traffic can assist in developing a robust model that will effectively recognize and deter possible denial of service or infiltration attacks.

## 1.2 Background Analysis

In recent years, technology has seen tremendous leaps in terms of availability, scalability, and adoptability both by companies and individuals alike. Technology has become an important part of our everyday lives and is also what fuels the expansion and profitability of most technology exposed companies.

With this exponential growth of technology, hacker groups and malicious individuals have found various ways to exploit the tools that companies use to their advantage to gain access to their networks or otherwise disrupt their normal operations. Those attacks can be devastating for a company's infrastructure and can lead to loss of money but also can lead to clients' personal data being stolen and potentially sold to malicious 3$^{rd}$ parties.

According to the European Union Agency for Cybersecurity ("ENISA"), the number of Denial-of-Service attacks have increased by 241% in 3Q 2019 when compared to the same period in 2018. This indicated that while companies aim to increase the level of Cybersecurity they implement, more and more attacks are being deployed every day that have the potential to cause harm. One recent example is the case of six major U.S. banks that were a target of a major wave of Denial-of-service attacks in 2012. According to CNN, no data has been stolen or otherwise tampered with during these attacks, however, all six banks have suffered a significant downtime in their services which means that, other businesses as well as individuals that use one of those banks to conduct their business transactions, have suffered a delay which could sometimes be catastrophic.

This is only one example of the millions of attacks happening every year. More research and development on sophisticated tools that could potentially mitigate or prevent all together such attacks are needed.

## 1.3 Need and Opportunity

As previously mentioned, the exponential growth of technology has also fueled the increase of different cyber-attacks that aim to disrupt the normal flow of operations. Such attacks have the potential to cause loss of revenue for businesses and even large-scale data loss and leaks.

To that end, not enough research has been conducted thus far on how the freely available data can fuel independent research that aims to limit or even stop such attacks from happening. Most available research papers use either closed source data to run their algorithms or have focused on how an algorithm can be used to predict such attacks using only one model. Due to the very private nature of network traffic logs, most companies tend to keep those records private and not release them to the public. Having no real-time network traffic logs available has led researchers and enthusiasts to generating simulated data based on lab-made networks which could have various disadvantages.
This paper will aim to explore the quality and usability of said data and how well they perform with different Machine Learning algorithms.

## 1.4 Research Questions

With the scarce nature of valid network traffic logs and the availability of simulated traffic instead, a few questions arise regarding their quality. This paper will aim to explore some of those questions and maybe provide some high-level answers that might help spark future research.

Due to the complicated and highly volatile nature of the field of Intrusion detection, we will try and answer the following:

1. Are the available open-source datasets enough to build a robust model to predict different kinds of attacks?
2. Are the extracted features of malicious traffic correlated to the features of benign traffic?
3. How can different Machine Learning algorithms improve performance and accuracy of the intrusion prevention predictions?
4. Do the results suggest that a passive detection approach (where the prediction is logged) is better than an active approach (where an action is taken based on the output)?

With the above questions we can hopefully provide a base framework for future research that can develop and hopefully produce a concrete model/understanding on how we can better combat network intrusion attempts.

## 1.5 Research plan and Thesis Outline

This thesis aims to use simulated datasets provided by the Canadian Institute of Cybersecurity (University of New Brunswick) such as their NSL-KDD and IDS-2017 datasets. Both those datasets provide network traffic information for benign traffic as well as various attacks.

The Canadian Institute of Cybersecurity datasets provide the network flows into different PCAP files as well as csv files extracted by a featured extraction tool called CICFlowMeter. This thesis will use the network extracted features as provided by CIC and will not in any way try to differentiate or analyze the effect of feature extraction in a models' accuracy score.

Both above datasets will be used as benchmarks for all open-source available datasets and will help us identify their strengths and weaknesses and plan for developing a concrete Intrusion Detection system.

# Literature Review

## 2.1 Dataset Availability and Review

The concept on Intrusion Detection and prevention has troubled scientists and network enthusiasts since the inception of the Internet. Business owners, researchers and "home-lab" owners alike, have been looking for ways to combat the exponential growth of Intrusion attempts on their systems. The field of Intrusion Detection has seen great strides over the last couple years with the rise of Machine Learning and a lot more research has been conducted. Cybersecurity experts expect the total cybercrime cost to increase by 15% per year over the next 4 years, up to nearly a total of $11 trillion US dollars by 2025 (Steve Morgan, 2020). With the potential of such great loss, it is clear that the Computer Science, Cryptography and Cyber Security fields could greatly benefit from the availability of extensive research and data that have the potential to spark innovations in the defense against such attacks.

Developing a model to identify malicious attempts accurately and consistently on a network is unfortunately a minor part of the entire problem. Markus Ring et al (2019) mention that "*Unfortunately, there are not too many representative data sets around… The community is working on this problem as several intrusion detection data sets have been published over the last years*". They also conclude that "*The community is aware of the importance of realistic network-based data, and this survey shows that there are many sources for such data (data sets, data repositories, and traffic generators)*". These statements, however, fail to recognize the interference that lab-generated and random-generated datasets can introduce into an Intrusion

Detection System. Simulated data are a great way of providing a baseline benchmark of the performance of a given model, or even being able to verify and prove that a given method could have the potential to detect and deter attacks, but logic suggest that they will probably not be the best fit for a production ready model. Roberto Magan-Carrion et al (2020), correctly point out that the reason no perfect dataset exists is that: a) new attacks are introduced fairly frequently and b) no one dataset can provide such a wide range as to capture the entire arsenal of attacks that malicious individuals or groups can unleash. With that being said, as previous authors point out, the currently existing datasets can sometimes be a great entry point into the Intrusion Detection world and all of them have their unique use-cases.

## 2.2 Currently Available Datasets

As previously mentioned, network traffic datasets have a significantly limited availability but there are however a few that have been previously used in evaluating Machine Learning models. Previous studies have extensively used the NSL-KDD and KDD Cup 1999 datasets for the development and evaluation of Intrusion Detection systems. S. Revanthi and Dr. A. Malathi (2013), highlighted that their statistical analysis identified major issues in the KDD Cup 1999 data set which affect the performance and estimation capability of various models. To resolve the issues of the KDD Cup 1999 data set, M. Tavallaee, E. Bagheri, W. Lu, and A. Ghorbani (2009), proposed a new dataset called NSL-KDD. The authors mention that the newly created dataset, - which is a re-shuffled and re-structured based on the KDD Cup 1999 dataset – does not suffer from all the problems that its predecessor did. Tavallaee, E. Bagheri, W. Lu, and A. Ghorbani (2009), claim that the NSL-KDD data set has no redundant records and that the test and train sets do not contain any shared records that might force a classifier to be biased towards one kind of family of attacks. The authors have also supported the claim of J. McHugh (2000), that, these kinds of datasets do not represent actual real-life networks and that deriving a simulated dataset that can accurately resemble a real-life network might not be possible at all. To that end, the need for more data sets have emerged and multiple researchers and companies have tried to create the "ultimate" data set.

Due to the limited availability of data sets for Intrusion Detection, Iman Sharafaldin, Arash Habibi Lashkari and Ali A. Ghorbani (2018), have proposed another open-source Intrusion Detection data set named CICIDS2017. The CIC IDS 2017 is a much more recent dataset when compared to NSL-KDD and KDD Cup 99 and it is aimed to be a data set that can help identify more kinds of attacks on a network when compared to older ones. The authors have created a testbed network that they then used to perform various attacks on while capturing the traffic using various tools. Iman Sharafaldin, Arash Habibi Lashkari and Ali A. Ghorbani (2018), mention in their paper that the attack schedule was split between 5 days of the week in the below manner:

*Table 1. List of Attacks*

| Days | Labels |
|---|---|
| Monday | Benign |
| Tuesday | Brute Force, SFTP, SSH |
| Wednesday | DoS, Heartbleed, SlowLoris, SlowHTTPTest, Hulk and GoldenEye |
| Thursday | Web and Inflitration, Web Brute Force, XSS, SQL Injection, Inflitration Dropbox Download and Cool Disk |
| Friday | DDoS. Botnet ARES, Portscans |

By using the above techniques to attack a network while capturing all traffic to and from it, the authors have captured a well-rounded understanding of what network attacks would look like from a traffic perspective. Abdullah Alsaeedi, Mohammad Zubair Khan (2019), in their recent paper, highlighted a few shortcomings of this new data set. According to the authors, the CIC IDS 2017 data set is significantly larger than other data sets and there is no separate test set from the training set. The two previously mentioned negatives, however, do cancel out in our opinion. Having a large data set is not always a negative when it comes to model performance and when it does become a problem, the effects are most of the times negligible. One good way to overcome such a problem is to use the k-fold Cross-Validation approach so that the model can effectively produce reliable accuracy and precision scores on one large data set. There are various other ways that one can use to split a data set into a training and test set so that they can be used for training and testing a model, however, due to the imbalanced nature of the problem, this can be very risky. The last disadvantage that Abdullah Alsaeedi, Mohammad Zubair Khan (2019) mention regarding the CIC IDS 2017 data set, is the low availability of papers that have used this data set to run models. The current low availability of studies focusing on the CIC IDS 2017 data set is by no means an indication of its performance and we should expect more scientists using this data set as a means of evaluating models. To that end, this paper will use both the NSL-KDD and CIC IDS 2017 datasets to evaluate the performance of plain machine learning models but also compare the benchmarking capabilities of the two freely available data sets.

## 2.3 Intrusion Detection Models

There are at least two different methodologies when attempting to create an Intrusion Detection System. Signature-based detection, where the defensive mechanism uses already known attacks to build a database and protect a system/network using that database, and Anomaly-based detection where machine learning is used to identify outliers in the data and tries to distinguish between normal and anomalous traffic. Many studies have been conducted in order to identify which model works best for intrusion detection and we now have various different tools in our arsenal to deter such attacks. Most of the tools however are using a signature-based approach to detect anomalous traffic for which SolarWinds N-Able (2021) says that *"...this potential disadvantage is also what makes anomaly-based intrusion detection able to detect zero-day exploits signature-based detection cannot"*. The ability to detect previously unknown attacks is a huge benefit to any organization/individual which aims to keep their data safe. Signature-based systems rely on the fact that an attack has been previously recorded and that we now have enough information to deter future attacks, where as an anomaly-based model will hopefully be able to deter an attack, it has not seen before.

Manjula C. Belavagi and Balachandra Muniyal (2016), run different Machine Learning classifiers on the NSL-KDD data set and have concluded that "…*the Random Forest classifier outperforms other classifiers for the considered data-set and parameters. It has the accuracy of 99%. The work can be extended by considering the classifiers for multiclass classification and considering only the important attributes for the intrusion detection.*". The authors have used the following four classifiers in their paper:

1) Logistic Regression
2) Support Vector Machine
3) Gaussian Naïve Bayes
4) Random Forest

The Random Forest model seems to be pairing great with the NSL-KDD data set which could probably be attributed to its ensemble nature and the fact that Intrusion Detection data sets are quite large. There is however a major drawback when using the Random Forest for Intrusion Detection. As correctly noted by Saurabh Kumar et al (2019), the Random Forest classifier is extremely time-consuming and expensive to construct. The authors also explain how a Random Forest model would require much more computational resources to keep running which makes it not ideal for an Intrusion Detection System.

# Research Methodology

## 3.1 Introduction

As clearly stated by previous researchers, the need for better Intrusion Detection data sets and models is dire. Right now, there is an extremely limited supply of data sets that research has been conducted on and therefore limits our capability of reliably benchmarking Machine learning models and training classifiers that could help in the battle against intrusion attempts.

Some of the major problems that have been discussed so far are:
- Limited availability of data sets.
- The limited available data sets provide a simulated view of the problem which can distort results
- Increased complexity of models which requires increased resources and a significantly longer inference time.

This paper will aim to compare two of the most well-known data sets currently available and by running multiple models on both, to try and provide a clear understanding of what makes an Intrusion Detection data set ideal for the task.

## 3.2 Collection of Data

To conduct our research and answer the questions we set previously, we are going to use two of the largest and well-known data sets currently available to the public. The two data sets are the NSL-KDD data set and the CIC IDS 2017 data set. The two data sets aim to provide low-level data of network traffic communication by including traffic that is considered as normal (non-malicious) and malicious.

*Table 2. Comparison of IDS Datasets

| Dataset | Year of Inception | No of Observations | No of Fields | Classes |
|---------|-------------------|--------------------|--------------|---------|
| CIC IDS 2017* | 2017 | 225,745 | 79 | Benign DDoS |
| NSL-KDD | 2009 | 125,973 | 42 | Anomaly Normal |

*Only data from the Friday file.

The CIC IDS 2017 dataset provides data in a comma-separated format (CSV), while the NSL-KDD dataset provides arff files which are native to the WEKA package. Both formats can be read by almost all programming languages which allows us to run virtually any classifier we would want.

We have chosen those two specific datasets for multiple reasons:
1. Significant age difference between the two (2009 – 2017) which might indicate that the latter one will be able to identify newer threats.
2. Significantly different observation and field sizes. This might help answer the question whether a larger dataset is better for Intrusion Detection.
3. Lack of research based on the CIC IDS 2017 dataset.

Our of the three main reasons, number three is an important one. While older datasets such as the NSL-KDD and KDD Cup 99 have been used multiple times in research papers, the CIC IDS 2017 have only been used a handful of times. Frequent testing of new and possibly improved datasets is a vital part in Intrusion Detection. The tools that attackers have are advancing exponentially, with more and more ways of intrusion being discovered. Due to that, the field of Cyber-Security needs to perform frequent testing and help the datasets grow so that they can detect and deter more "modern" attacks.

## 3.3 Machine Learning Classifiers

Intrusion Detection is a time-sensitive task that if not handled on time, could have devastating results. To that end, having a model that can run inference on incoming data in real or close to real-time, is essential to deter said attacks. While this paper will not directly compare the inference time between models, it is still an incredibly significant part of the entire Intrusion Detection effort.

To compare the performance of our two datasets, we will run them through several different classification algorithms. All models will use the same settings across both datasets to ensure a clear view of how the actual dataset affects performance. T. Saranya, S. Sridevi, C. Deisy, Tran Duc Chung and M.K.A Ahamed Khan (2020), found that, the Random Forest classifier has outperformed all other tested classifiers by a significant margin with an Accuracy rate of 99.65%. There is however, no shortage of classifiers that can be a great fit for an Intrusion Detection system and this paper will aim to run most of the major classifier on the two datasets mentioned above.

### 3.3.1 J48 (C4. 5 Decision Tree)

The J48 model is a classifier that is native to the WEKA package written in Java. The J48 classifier is used to generate a C4.5 Decision tree. Decision Trees are used to generate a "decision" given a dataset. The trees are created based on nodes and assigned a probability of an event happening given the outcome of the "parent nodes". We expect that this classifier will be a great fit for an Intrusion Detection system due to its nature and the nature of network traffic.

### 3.3.2 Random Forest

The Random Forest deploys several different decision trees and applies them on sub-sections of a given dataset. The Random Forest classifier then combines the outcome of the decision trees to significantly increase the prediction accuracy.

### 3.3.3 Isolation Forest

The Isolation Forest classifier belongs in the family of anomaly-detection (or misuse detection) algorithms. Fei Tony Liu, Kai Ming Ting and Zhi-Hua Zhou (2008) are the original authors of the Isolation Forest classifier and in the paper state that: "*Since anomalies are 'few and different' and therefore they are more susceptible to isolation. In a data-induced random tree, partitioning of instances are repeated recursively until all instances are isolated.*". During their evaluation, the Isolation Forest algorithm appeared to outperform other classifiers such as the Random Forest in an Anomaly detection setting. Based on that, the Isolation Forest might be worth examining for Intrusion Detection.

### 3.3.4 AdaBoost.M1

The AdaBoost.M1 classifier belongs in the family of *Boosting* classifiers. Boosting classifiers try to create a stronger model that is based on a series of other weaker models. To achieve that, the base model is run multiple times with each iteration trying to resolve the errors of the previous one. The boosting algorithm runs this procedure multiple times until the dataset is fully explained by the model or until a predefined hard iteration limit is reached. Not many papers have used any boosting algorithms to tackle problems in the Intrusion Detection space and therefore, comparing datasets based on one might return results worth noting.

## 3.4 Methodology Summary

Different machine learning classifiers provide multiple different benefits to each application. For this paper we have selected two representative datasets and four different classifiers. By running all four classifiers on each dataset, we hope to answer the question of which dataset provides greater inside as to which traffic entries and malicious and which are not. As previously stated, we have selected two quite different datasets both in terms of age but also in terms of data quality, length, and number of features.

All classifier outputs will be rated based on their Accuracy and Precision scores both between them but also across our two datasets.

Note that, while this exercise will give us great insight as to which classifier performs best on each dataset, it is by no means a definitive result. More research will need to be conducted to identify low-level correlation between datasets as well as, what makes a dataset unique and ideal of a real-time, production-ready model

# Analysis of Research Findings

## 4.1 Datasets

As mentioned in previous sections, we will be using two of the currently most popular datasets available for Intrusion Detection. The two datasets are the CIC IDS 2017 and NSL-KDD. As we have seen, the two datasets differ in various ways which could lead to one of the two to stand out in terms of their predictive capabilities.

**\*Figure 1. Observations of Datasets**

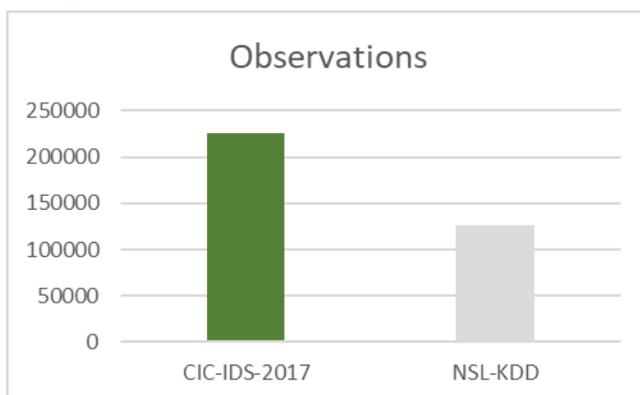

**\*Figure 2. Number of Features**

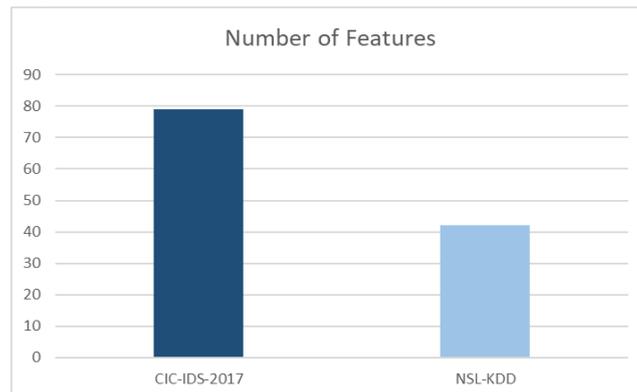

To that end, understanding the difference between the two is essential to gain a greater insight as to what constitutes a concrete Intrusion Detection dataset. Going forward, we will dive deeper into how the two datasets are constructed and what makes each one unique.

### 4.1.1 CIC-IDS-2017 Dataset

The CIC IDS 2017 dataset is a collection of datapoint from simulated attacks on a controlled network. It provides network features extracted from normal/benign traffic but also from various different attack types such as DDoS ("Distributed Denial Of Service"), PortScan, Infiltration and more. For this paper, the different kinds of attack were merged under the "anomaly" class to create a composite binary class and provide a dataset that more closely resembles NSL-KDD. This paper will not assess the capability of a model in cherry-picking out the difference between various attacks. The dataset provides two sets of data, the original PCAP files with the network traffic log and CSV files which are derived from running the PCAP files through CICFlowMeter, a feature extraction tool which extracts network features from PCAP files in plain text. The resultant 78 network features consist of various metrics of network traffic such as the duration of the flow of data, the port number that the communication was forwarded to, various metrics on the packet size and much more.

**\*Table 3. CIC-IDS-2017 Class Distribution**

| Class | Frequency | Percentage |
|---|---|---|
| Benign | 97718 | 43.28% |
| Malicious | 128027 | 56.71% |

Intrusion Detection is well known to be a very imbalanced problem, meaning that the occurrence of malicious network traffic happens at a significantly lower rate than normal traffic. While normal traffic is logged all the time, malicious traffic can only be logged during an attack therefore creating a gap in the availability of malicious datapoints when compared to normal datapoints. The CIC IDS 2017 dataset is portraying a naturally imbalanced problem as a highly balanced one. Being a research dataset, this is a requirement since an imbalanced class variable could lead to catastrophic model failures.

Due to the high amount of network features available, it is important to account for dimensionality in our models.

*Table 4. CIC-IDS-2017 Selected Features

| Feature | Weight |
|---|---|
| Total Length of Fwd Packets | 0.6510714 |
| Subflow Fwd Bytes | 0.6510714 |
| Average Packet Size | 0.5607741 |
| Total Length of Bwd Packets | 0.5423789 |
| Subflow Bwd Bytes | 0.5423789 |
| Bwd Packet Length Mean | 0.5419528 |
| Avg Bwd Segment Size | 0.5419528 |
| Fwd Header Length | 0.5391932 |
| Fwd Header Length 1 | 0.5391932 |
| Destination Port | 0.5377644 |

To accomplish that, we can employ the Information Gain criterion. Information gain is a feature selection tool that will help us identify which of our features hold the most information to describe our Class variable (Benign or Malicious traffic). The weights represent the rate at which each of the selected feature explains our Class variable and we as we have found, the features of Table 4 appear to explain the class variable with significant "accuracy".

*Table 5. CIC-IDS-2017 Selected Features Statistics

| Label | BENIGN | | | DDoS | | |
|---|---|---|---|---|---|---|
| Variable | N | Mean | SD | N | Mean | SD |
| Total.Length.of.Fwd.Packets | 97714 | 2128.6 | 4679.729 | 128027 | 31.909 | 11.957 |
| Subflow.Fwd.Bytes | 97714 | 2128.6 | 4679.729 | 128027 | 31.909 | 11.957 |
| Average.Packet.Size | 97714 | 249.601 | 388.137 | 128027 | 822.612 | 658.845 |
| Total.Length.of.Bwd.Packets | 97714 | 4109.172 | 59215.5 | 128027 | 7373.635 | 5584.659 |
| Subflow.Bwd.Bytes | 97714 | 4109.172 | 59215.5 | 128027 | 7373.635 | 5584.659 |
| Bwd.Packet.Length.Mean | 97714 | 116.901 | 257.698 | 128027 | 1481.026 | 1164.873 |
| Avg.Bwd.Segment.Size | 97714 | 116.901 | 257.698 | 128027 | 1481.026 | 1164.873 |
| Fwd.Header.Length | 97714 | 130.427 | 568.949 | 128027 | 97.096 | 38.291 |
| Fwd.Header.Length.1 | 97714 | 130.427 | 568.949 | 128027 | 97.096 | 38.291 |
| Destination.Port | 97714 | 20406.712 | 25829.239 | 128027 | 81.227 | 267.41 |

To better understand the differences between our two classes, it is best that we look at what "information" our selected features can offer. We can see that there are significant differences in the mean values between the two. We can conclude that malicious traffic has a much lower forward packet length and subflow when compared to benign traffic but on the other hand, it has a significantly higher backwards flowing packet length. Destination port gives us little information as to how the traffic changes between benign and malicious since it only indicates the actual destination of a packet. In this dataset, DDoS traffic seems to favor lower numbered ports while benign traffic favors much higher port ranges. This might be the case because this is a simulated dataset, and the authors of the dataset wanted a clear distinction between the two classes.

### 4.1.2 NSL-KDD

Similarly with the CIC IDS 2017 dataset, the NSL-KDD provides an insight into normal and anomalous traffic data. This dataset, however, provides its data mainly through *arff* files which are native to the WEKA Machine Learning package written in Java.

**\*Table 6. NSL-KDD Class Distribution**

| Class | Frequency | Percentage |
|---|---|---|
| *Benign* | 67343 | 53.45% |
| *Malicious* | 58630 | 46.54% |

The NSL-KDD dataset provides a significantly lower amount of datapoints when compared to CIC IDS 2017 but both datasets appear to provide a balanced view to the problem in question. Both Classes (normal and malicious), appear to have an almost equal amount of datapoints which is very helpful in the development of an unbiased model.

**\*Table 7. NSL-KDD Selected Features**

| Feature | Weight |
|---|---|
| src_bytes | 0.5653506 |
| service | 0.4654933 |
| dst_bytes | 0.4380953 |
| flag | 0.3600125 |
| diff_srv_rate | 0.3587808 |
| same_srv_rate | 0.3533069 |
| dst_host_srv_count | 0.3294796 |
| dst_host_same_srv_rate | 0.3031884 |
| dst_host_diff_srv_rate | 0.2841864 |
| dst_host_serror_rate | 0.2807353 |

The features of NSL-KDD also seem to explain the Class variable considerably well, however, when compared to CIC IDS 2017, the differences are significant. The top 10 selected features of the CIC IDS 2017 explained the class variable with a weight higher than 0.5 whereas the NSL-KDD ones seem to drop below 0.5 by a significant margin. This could in theory be a major advantage of the CIC IDS 2017 but this will only become evident once both datasets have been used into Machine Learning models to detect the performance difference these could yield in a model.

**\*Table 8. NSL-KDD Selected Features Statistics**

| Label | normal | | | anomaly | | |
|---|---|---|---|---|---|---|
| Variable | N | Mean | SD | N | Mean | SD |
| src bytes | 67343 | 13133.279 | 418113.134 | 58630 | 82820.141 | 8593024.6 |
| dst bytes | 67343 | 4329.685 | 65462.818 | 58630 | 37524.482 | 5893990.938 |
| diff srv rate | 67343 | 0.029 | 0.146 | 58630 | 0.102 | 0.206 |
| same srv rate | 67343 | 0.969 | 0.144 | 58630 | 0.307 | 0.396 |
| dst host srv count | 67343 | 190.286 | 92.608 | 58630 | 29.929 | 52.289 |
| dst host same srv rate | 67343 | 0.812 | 0.324 | 58630 | 0.187 | 0.322 |
| dst host diff srv rate | 67343 | 0.04 | 0.129 | 58630 | 0.132 | 0.231 |
| dst host serror rate | 67343 | 0.014 | 0.092 | 58630 | 0.595 | 0.484 |

With a clear understanding of how the two datasets work and what they consist of, it is important to benchmark the two against various models to understand how each classifier perform under different models.

As stated, this paper will focus on four classifiers to perform its comparison:
1. J48
2. Random Forest
3. Isolation Forest
4. AdaBoost.M1 (using a Decision Stump as its weak classifier)

*4.2.1 J48*
Previous research has found great success when using the J48 classifier on intrusion data. To that end, we run the J48 classifier on our two datasets using the WEKA package.

**Table 9. J48 Classifier Comparison**

| | **CIC IDS 2017** | | | *NSL-KDD* | | |
|---|---|---|---|---|---|---|
| | Benign | Malicious | | | Benign | Malicious |
| *Benign* | 29309 | 5 | | *Benign* | 9448 | 263 |
| *Malicious* | 6 | 38402 | | *Malicious* | 3900 | 8933 |

*Figure 3. J48 Performance Metrics

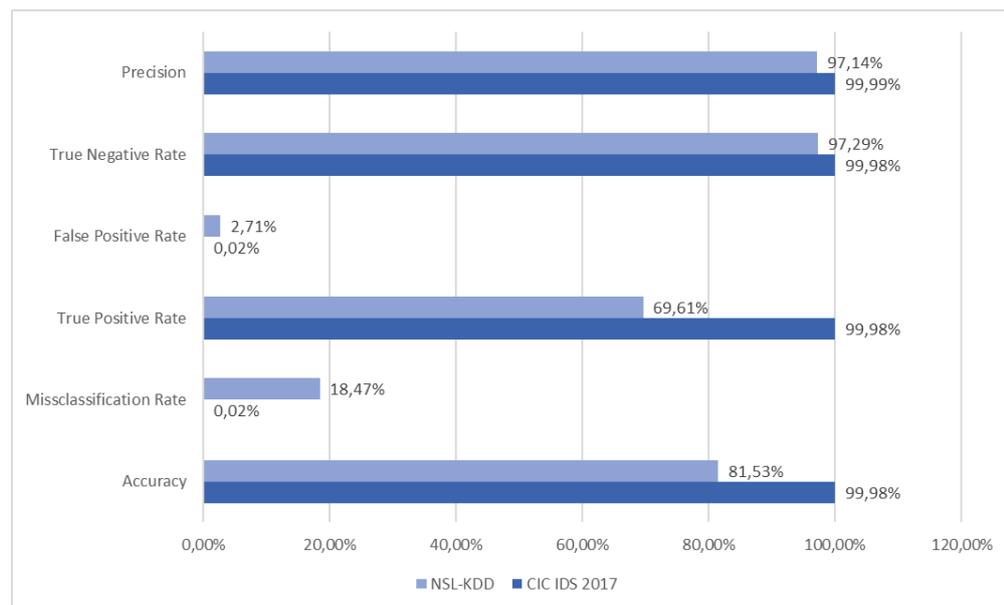

### 4.2.2 Random Forest

As mentioned in Chapter 4. Literature Review, most previous researchers have identified that the Random Forest Classifier is the best one can choose. Previous researchers however have focused on the NSL-KDD and KDD Cup 99 datasets. Running the Random Forest on a newer dataset such as the CIC IDS 2017, will help us identify whether the classifier is still the rightful king.

Table 10. Random Forest Classifier Comparison

| | CIC IDS 2017 | | | NSL-KDD | | |
|---|---|---|---|---|---|---|
| | Benign | Malicious | | | Benign | Malicious |
| *Benign* | 29303 | 0 | | *Benign* | 67317 | 26 |
| *Malicious* | 4 | 38416 | | *Malicious* | 685 | 57945 |

*Tree depth reduced to 5 to avoid overfitting

*Figure 4. Random Forest Performance Metrics

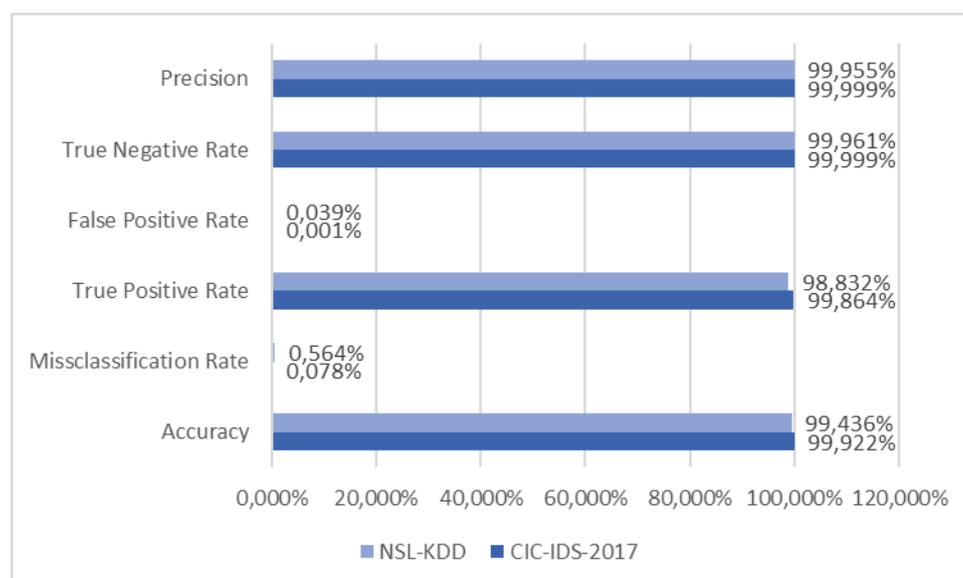

We have been able to confirm the results of previous studies and identify that this classifier performs exceptionally well on both our datasets. The classifier managed to achieve a 99.9% accuracy and precision scores on both datasets, providing an exceptionally low False Positive rate.

### 4.2.3 Isolation Forest
The Isolation Forest is an algorithm that is part of this misuse detection family of models. Intrusion Detection is at its core a detection of anomalies and would be ideal if we could examine the performance of one such model.

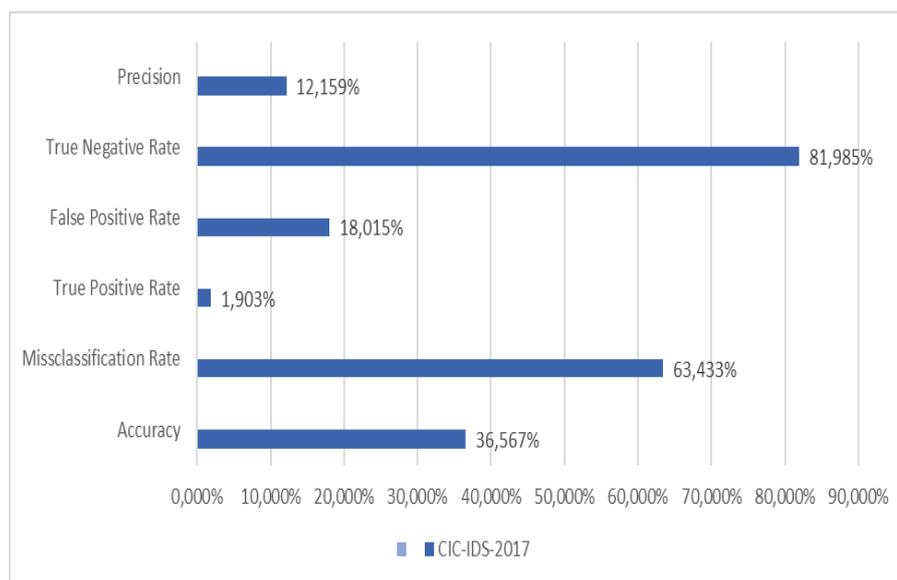

*Figure 5. Isolation Forest Performance Metrics

Unfortunately, the NSL-KDD dataset as provided by the authors, cannot be loaded, and used by the Isolation Forest classifier used in the WEKA package. Digging further into this problem, this issue seems to arise from the declaration of fields in the NSL-KDD dataset.

The Isolation Forest classifier on the CIC IDS 2017 dataset however, seems to perform rather poorly. Especially when compared to other Decision Tree based classifiers. This low performance could be attributed to the fact that this is a simulated dataset that has been made balanced for the sake of developing theoretical models. Other Anomaly detection algorithms should be used and implemented by future researchers since Intrusion Detection is a great example of what an Anomaly Detection classifier should be able to do.

### 4.2.4 AdaBoost.M1

The final model we will try to explain on our two datasets is a gradient boosting model. The AdaBoost.M1 model tries to create a strong classifier based on the outcome and correction of errors of multiple weaker models, in our case a Decision stump model.

**Table 11. AdaBoost.M1 Classifier Comparison**

| | CIC IDS 2017 | | | NSL-KDD | |
|---|---|---|---|---|---|
| | Benign | Malicious | | Benign | Malicious |
| *Benign* | 29273 | 30 | *Benign* | 19906 | 261 |
| *Malicious* | 52 | 38368 | *Malicious* | 808 | 16817 |

\*Figure 6. AdaBoost.M1 Performance Metrics

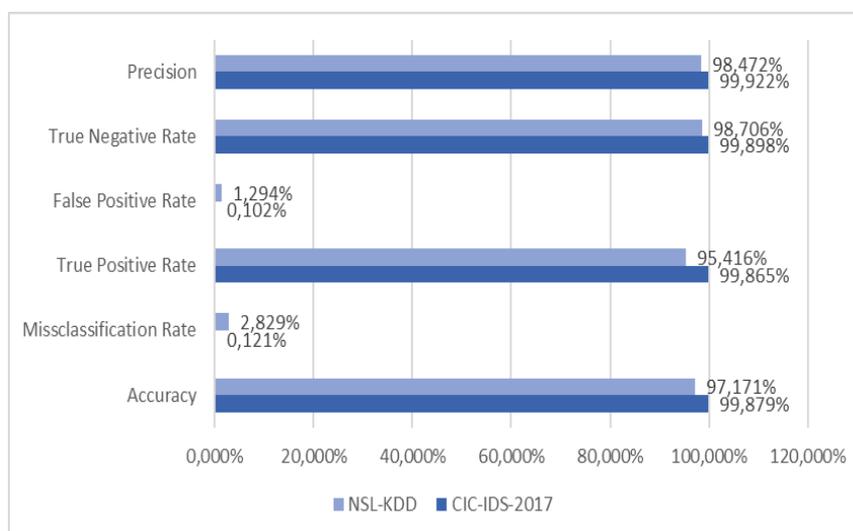

As per our previous models, the AdaBoost.M1 model seems to perform exceptionally well, reaching accuracy scores of 99.9% for the CIC IDS 2017 dataset and 97.2% for the NSL-KDD. The accuracy ratio of the classifier when used on CIC IDS 2017 is marginally better but close to NSL-KDD.

We have identified a major drawback however when it comes to using the AdaBoost.M1 classifier. The classifier was considerably slower at generating predictions when compared to previous models and that might indicate that it is not suited for a production environment where inference speed is essential.

# Conclusions

As our research suggests, both freely available datasets perform quite well using certain Machine Learning algorithms. The training times are high given the amount of data; however, the prediction times are relatively low which indicates that they could very well be used in a production ready model. There are although some caveats. Both datasets display a balanced approached to a wildly imbalanced problem. Both being simulated data; they assume that the distribution of normal and anomaly data points is equal which in a real-life scenario they wouldn't be. They also offer a limited variety of class attributed such as "anomaly" and "normal" but a production ready model should be able to identify and distinguish between different kinds of attacks and be able to "protect" the network from all of them while taking certain actions to mitigate said attacks. The older dataset, NSL-KDD, does not provide a distinction between various kinds of attacks but rather provides details as to whether a datapoint is normal or malicious. CIC IDS 2017 on the other hand, distinguishes between different attacks and it would be beneficial if future research tries to identify whether such distinction would benefit a model.

We can clearly see that the CIC IDS 2017 dataset seems to perform better than the NSL-KDD dataset in certain situations. This could be attributed to a few key points. As we have previously seen, the CIC IDS 2017 dataset provides a significantly higher observation number and much more features. This could mean that, a researcher can select more features that explain the Class variable with a high weight. This would lead to a model that can better extrapolate the effect of a given variable on the class and easily distinguish between what is normal traffic and what is malicious traffic. The CIC IDS 2017 dataset is also a significantly newer collection of data which is a significant advantage. As stated, intrusion attempts grow in size and sophistication every year and for that, having an updated dataset can significantly help in the effort of battling such attacks.

 We can, however, safely assume that both above datasets were created with an academic purpose in mind and are not meant to be used in a production setting despite their high accuracy scores. A lot more research is required to identify and explain how a model can achieve insanely fast prediction speeds and accuracy scores so that it will not interfere with the speed of a network which is critical for a business. Rule-based detection on the other hand handles this very well. It can:
1. Drop traffic on certain ports
2. Drop traffic from hosts with no recognized keys
3. Drop traffic coming from certain domains with a high chance of being malicious
 and much more which still render them one of the top-choices in the intrusion detection space.

To conclude, it is evident that the availability of network data is scarce and that we need better datasets to help us create more robust models that can reflect real-life scenarios of attacks happening on real networks. As we have seen, both datasets perform exceptionally well in certain situations and from the evidence that this paper provides, we can safely assume that both datasets will be great tools to kickstart the research needed to reduce or even eliminate network attacks. We propose that, an open-source dataset is created that would always keep up to date datapoints of network attacks and normal traffic so that future researchers can train models and identify new combination of classifiers that can in real-time deter attacks.

# References

Iman Sharafaldin, Arash Habibi Lashkari, and Ali A. Ghorbani (2018) Toward Generating a New Intrusion Detection Dataset and Intrusion Traffic Characterization, 4th International Conference on Information Systems Security and Privacy (ICISSP), Portugal, January

M. Tavallaee, E. Bagheri, W. Lu, and A. Ghorbani (2009) A Detailed Analysis of the KDD CUP 99 Data Set, Submitted to Second IEEE Symposium on Computational Intelligence for Security and Defense Applications (CISDA)

Steve Morgan (2020, November 13), *Cybercrime to cost the world $10.5 trillion annually by 2025, https://cybersecurityventures.com/hackerpocalypse-cybercrime-report-2016/*

Markus Ring, Sarah Wunderlich, Deniz Scheuring, Dieter Landes, Andreas Hotho (2019), *A Summary of Network-based Intrusion Detection Data Sets,* University of Applied Sciences, Coburg Germany

J. McHugh (2000), *Testing Intrusion Detection Systems: a Critique of the 1998 and 1999 DARPA Intrusion Detection System Evaluations as performed by Lincoln Laboratory,* ACM Transaction on Information and System Security.

Abdullah Alsaeedi, Mohammad Zubair Khan (2019), *Performance Analysis of Network Intrusion Detection System using Machine Learning,* Department of Computer Science, Taibah University, Madinah, KSA.

N-Able (2021, March 21) *Intrusion Detection System (IDS): Signature vs. Anomaly-Based*, https://www.n-able.com/blog/intrusion-detection-system

Manual C. Belavagi, Balachandra Muniyal (2016), *Performance Evaluation of Supervised Machine Learning Algorithms of Intrusion Detection,* Manipal Institute of Technology, Manipal, India

T. Saranya, S. Sridevi, C. Deisy, Tran Duc Chung and M.K.A Ahamed Khan (2020), *Performance Analysis of Machine Learning Algorithms in Intrusion Detection System: A Review,* Department of Information Technology, Thiagarajar College of Engineering, Madurai, India

Fei Tony Liu, Kai Ming Ting and Zhi-Hua Zhou (2008), *Isolation Forest,* Gippsland School of Information Technology Monash University, Victoria, Australia and National Key Laboratory
for Novel Software Technology Nanjing University, Nanjing, China